\documentclass[12pt,preprint]{aastex}
\usepackage{emulateapj5}
\usepackage{apjfonts}
\usepackage{epsf}

\journalinfo{The Astrophysical Journal Letters, 607:L101-L111, 2004 May 1}

\def\cj{{\cal J}}

\def\pf{\langle f\rangle}

\begin{document}

\title{Hot Disk Corona and Magnetic Turbulence in Radio-Quiet Active Galactic Nuclei:
Observational Constraints}

\author{Jian-Min Wang\altaffilmark{1,2},
        Ken-Ya Watarai\altaffilmark{1}
        and Shin Mineshige\altaffilmark{1}
        }
        
\altaffiltext{1}{Yukawa Institute for Theoretical Physics, Kyoto University,
Kyoto 606-852, Japan}
                      
\altaffiltext{2}{Laboratory for High Energy Astrophysics, Institute of
High Energy Physics, Chinese Academy of Science, Beijing 100039, P.~R. China.}

\slugcomment{Received 2004 March 3; accepted 2004 April 16; published 2004 April 20}
\shorttitle{Hot Disk Corona and Magnetic Turbulence}
\shortauthors{Wang, Watarai \& Mineshige}

\begin{abstract}
We compile a sample consisting of 56 radio-quiet active galactic nuclei 
so as to investigate statistical properties of hot corona of accretion disks from 
{\em ASCA} observations. The black-hole masses in the sample are estimated via 
several popular methods 
and the bolometric luminosities from the multi-wavelength continuum. This allows 
us to estimate the Eddington ratio (${\cal E}\equiv L_{\rm Bol}/L_{\rm Edd}$)
so that the undergoing physical processes can be tested via hard X-ray data.  
We find a strong correlation between 
${\cal F}_{\rm X}\equiv L_{\rm 2-10keV}/L_{\rm Bol}$ and ${\cal E}$ as 
${\cal F}_{\rm X}\propto {\cal E}^{-0.64}$ with a multivariate regression. This 
indicates that the release of gravitational energy in the hot corona is controlled 
by the Eddington ratio. On the other hand, the correlation between the hard X-ray 
spectral index ($\Gamma$) and ${\cal E}$ depends critically on the types of objects: 
$\Gamma$ is nearly constant ($\Gamma \propto {\cal E}^0$) in broad-line Seyfert 1's 
(BLS1s), whereas $\Gamma \propto \log {\cal E}^{0.18}$ in narrow-line Seyfert 1's 
(NLS1s), although 
not very significant. We can set constraints on the forms of magnetic stress tensor 
on the condition that ${\cal F}_{\rm X}$ is proportional to the fraction 
$f$ of gravitational energy dissipated in the hot corona and that $f$ is proportional 
to magnetic energy density in the disk. We find that the shear stress tensor
$t_{r\phi}\propto P_{\rm gas}$ is favored by the 
correlation in the present sample, where $P_{\rm gas}$ is the gas pressure.   

\end{abstract}


\section{Introduction}
Accretion onto a supermassive black hole is generally regarded to be
powering active galactic nuclei (AGNs). 
For a steady Keplerian accretion disk around a black hole with mass $M_{\rm BH}$,
the released energy via viscosity dissipation is given by
\begin{equation}
Q_{\rm vis}=8.6\times 10^{25}\dot{m}m_{\rm BH}^{-1}r^{-3}\cj(r),~~~~~
{\rm (erg~s^{-1}~cm^{-2})}
\end{equation}
where $\dot{m}=\dot{M}c^2/L_{\rm Edd}$, the Eddington luminosity
$L_{\rm Edd}=4\pi c GM_{\rm BH}/\kappa_{\rm es}$, 
$\kappa_{\rm es}=0.34$, $r=Rc^2/2GM_{\rm BH}$,
$\cj(r)=1-(3/r)^{1/2}$ and $m_{\rm BH}=M_{\rm BH}/M_{\odot}$
(Shakura \& Sunyaev 1973). It is important to note that $Q_{\rm vis}$ is independent of 
the specific mechanism of viscosity stress. In the regime of the standard accretion disk 
around a massive black hole, the disks can not produce X-ray emission. Models of the 
accretion disks with hot corona have been suggested in order to explain the observed 
X-ray emission (Liang \& Price 1977). In a popular model, a fraction $f$ of $Q_{\rm vis}$ 
is assumed to be released in the hot corona and the left ($1-f$) in cold disk, but the 
transportation of angular momentum is mainly taking place in cold disk (Haardt \& Maraschi 
1991; Svensson \& Zdziarski 1994, Kawaguchi et al. 2001). However, the factor $f$ remains
open since we poorly understand how the process happens in the disk.

The important roles of the magnetic field turbulence  have been realized in transportation 
of angular momentum (Balbus \& Hawley 1991, Tout \& Pringle 1992, Turner et al. 2002, 2003) 
and formation of the hot corona (Galeev et al. 1979, Stella \& Rosner 1984, Merloni \& Fabian 
2002, Liu et al. 2002, Turner 2004, Kuncic \& Bicknell 2004). The Maxwell stress is capable 
of transporting the angular momentum in the disk (Balbus 2003). Numerical simulations show 
the actual physical processes inside the disks are very complicated (Balbus \& Hawley 1991, 
Stone et al. 1996; Blaes 2002; Blaes 
\& Socrates 2003; Machida \& Matsumoto 2003; Igumenshchev et al. 2003; Sano et al. 2003, 
Kato et al. 2004). We noted that these simulations are based on
the assumption of radiative inefficiency, and can not be directly applied to explain 
observations. The seminal paper by Galeev et al. (1979) showed that the 
gravitational energy will be mainly released in the hot corona since the strong buoyancy 
and magnetic field reconnection inevitably lead to the formation of the hot corona above 
cold disk (Stella \& Rosner 1984). The formation of the hot corona 
is most likely naturally related with the transportation of angular momentum, the
factor $f$ can be thus obtained if the magnetic stress and energy transportation are assumed 
(Merloni \& Fabian 2002). This may lend us a possible opportunity 
to test the working magnetic stress from hard X-ray observations. There is growing evidence 
in the Galactic black hole candidates for that the hot corona becomes weak when $\dot{m}$ 
increases (see a review of McClintock \& Remillard 2004). However it is elusive how the hot 
corona changes with $\dot{m}$ in AGN disks. Two problems remain open: 1) what is the proper 
magnetic field stress? 2) how to connect the two processes of angular momentum transportation 
and the energy release
in the hot corona?

In this {\em Letter}, we look into the hard X-ray data of radio-quiet AGNs
to see $\dot{m}$-dependence of hard X-rays and consider the roles of magnetic turbulence 
in the disks. We found a strong correlation between the factor $f$ and the Eddington ratio, 
implying that the energy release is driven by the Eddington ratio and the importances of 
magnetic turbulence. 

\section{The Sample and Correlations}
Table 1 gives the data 
of the present sample, which is composed of 56 objects
consisting of 29 narrow line 
Seyfert 1 galaxies (NLS1s) and 27 broad Seyfert 1 galaxies (BLS1s).  
Col (1) is the name of the objects, Col (2) the 2-10keV luminosity 
from {\em ASCA} observations in units of erg~s$^{-1}$, Col (3) the photon index between 
2-10 keV, Col (4) the bolometric luminosity in units of erg~s$^{-1}$, Col (5) the black hole 
mass in the units of solar mass. Col (6) gives some notes on the objects, indicating the 
references of X-ray luminosity and photon index as well as the references or methods of the 
estimations of the black hole mass, bolometric luminosity.
The black hole masses are estimated by

\begin{center}
\footnotesize
\centerline{\sc Table 1 The Sample}
\vskip 0.2cm
\begin{tabular}{llcccl}\hline \hline
Name &$\log L_{\rm X}$& $\Gamma$& $\log L_{\rm Bol}$& $\log m_{\rm BH}$& Notes\\ 
(1)  &(2)             &(3)      &(4)                &(5)    & (6) \\ \hline
Mrk 335        & 43.42&  1.94& 44.69&  6.69&T, 1   \\
IZw 1          & 43.73&  2.25& 45.47&  7.26&T, 2   \\
Ton S180       & 44.00&  2.46& 45.70&  6.91&T, 3   \\
F 9            & 44.26&  1.91& 45.23&  7.91&T, 1   \\
RX J0148-27    & 43.81&  2.06& 45.68&  6.97&T, 3   \\
Nab 0205+024   & 44.52&  2.27& 45.45&  7.86&T, 1   \\
Mrk 1040       & 42.83&  1.56& 44.53&  7.64&T, 1   \\
LB 1727        & 44.59&  1.56& 46.37&  8.47&T, Rev \\
3C 120         & 44.34&  1.89& 45.34&  7.42&T, 1   \\
Akn 120        & 44.07&  1.91& 44.91&  8.27&T, 1   \\
MCG+8-11-11    & 43.54&  1.56& 44.16&  7.24&T, Rev \\
1H0707-495     & 42.95&  2.27& 44.43&  6.30&T, Rev \\
NGC 3227       & 42.01&  1.61& 43.86&  6.76&T, 4   \\
RE 1034+396    & 43.05&  2.35& 44.52&  6.45&T, 3   \\
NGC 3516       & 43.43&  1.83& 44.29&  7.36&T, 1   \\
NGC 3783       & 43.25&  1.70& 44.41&  6.94&T, 1   \\
Mrk 42         & 42.54&  2.14& 43.91&  5.61&T, Rev \\
NGC 4051       & 41.56&  1.92& 43.56&  6.13&T, 1   \\
NGC 4151       & 43.01&  1.57& 43.73&  7.13&T, 1   \\
Mrk 766        & 43.08&  2.16& 44.23&  6.05&T, 3   \\
NGC 4593       & 43.06&  1.78& 44.09&  6.91&T, 1   \\
IRAS 13224-3809& 43.11&  1.97& 45.74&  6.75&T, Rev \\
MCG-6-30-15    & 43.11&  2.02& 43.59&  6.21&T, Rev \\
EXO 055620-3   & 43.97&  1.70& 44.37&  7.21&T, Rev \\
IC 4329A       & 43.06&  1.71& 44.78&  6.77&T, 1   \\
Mrk 279        & 43.99&  2.04& 44.55&  7.83&T, Rev \\
PG 1404+226    & 43.49&  2.07& 45.29&  6.76&T, Rev \\
NGC 5548       & 43.76&  1.79& 44.83&  8.03&T, 1   \\
Mrk 478        & 43.83&  2.06& 45.42&  7.27&T, Rev \\
Mrk 841        & 43.83&  1.77& 45.84&  8.10&T, 1   \\
Mrk 290        & 43.53&  1.68& 44.35&  7.04&T, Rev \\
3C 390.3       & 44.41&  1.63& 44.88&  8.55&T, 1   \\
Mrk 509        & 44.38&  1.82& 45.03&  7.86&T, 1   \\
Akn 564    & 43.38$^*$&  2.70& 44.47&  6.04&T, Rev \\
MCG-2-58-22    & 44.43&  1.73& 45.30&  8.48&T, Rev \\
E 0015+162     & 45.10&  1.99& 45.31&  8.31&R, 5   \\
PHL 909        & 44.24&  1.11& 45.47&  8.90&R, 1   \\
HE 1029-1401   & 44.44&  1.81& 46.03&  9.08&R, 1   \\
PG 1114+445    & 44.01&  1.71& 45.65&  8.41&R, 2   \\
PG 1116+215    & 44.47&  2.09& 46.22&  8.50&R, 2   \\
PG 1211+143    & 43.70&  2.06& 45.89&  7.88&R, 2   \\
PG 1216+069    & 44.70&  1.57& 46.45&  9.17&R, 2   \\
PG 1404+226    & 42.95&  1.77& 45.27&  6.74&R, Rev, 6\\
PG 1416-129    & 44.48&  1.78& 45.90&  8.50&R, 2    \\
PHL 1092       & 44.10&  1.99& 46.07&  7.61&V, Rev  \\
RX J0439-45    & 44.04&  2.25& 45.66&  7.12&V, Rev  \\
PKS 0558-504   & 44.67&  1.81& 45.80&  7.56&V, Rev  \\
PG 1244+026    & 42.10&  2.35& 44.62&  6.29&V, 2    \\
PG 1543+489    & 44.16&  2.46& 46.11&  7.79&V, Rev  \\
IRAS 1702+454  & 43.74&  2.20& 44.74&  6.54&V, Rev  \\
Mrk 507        & 42.62&  1.61& 44.54&  6.34&V, Rev  \\
IRAS 20181-224 & 43.87&  2.33& 45.43&  6.80&V, Rev,7\\
Mrk 142        & 43.18&  2.12& 44.55&  6.65&L, 3    \\
IRAS 13349+2438& 44.12&  2.31& 45.35&  7.74&L, 3    \\
Kaz 163        & 43.19&  1.92& 45.02&  7.11&L, Rev  \\
NGC 7469       & 43.26&  1.91& 45.28&  6.84&N, 1    \\ \hline
\end{tabular}
\vskip 0.2cm
\parbox{3.45in}
{\baselineskip 9pt
\indent
{\sc Notes/References:--}
T: Turner et al. (1999); R: Reeves \& Turner (2000); V: Vaughan et al. (1999);
L: Leighly (1999); N: Nandra et al. (2000); 
1: Woo \& Urry (2002); 
2: Vestergaard (2002); 
3: Grupe et al. (2004);
4: McLure \& Dunlop (2001);
5: \"Orndahl et al. (2003); 
6: Boroson \& Green (1992); 
7: Elizalde \& Steiner (1994);
Rev: reverberation relation from the magnitude $M_{\rm B}$ from V\'eron-Cetty \& V\'eron Catalog; 
* data is taken from Turner et al (2001).}
\end{center}
\normalsize

\noindent
1) the empirical reverberation relation with calibrations, 
$R_{\rm BLR}=30.2\left(L_{\rm 5100\AA}/10^{44}\right)^{0.66}$lt-days, 
where $L_{\rm 5100\AA}$ is the luminosity at 5100\AA (Vestergaard 2002); 
2) reverberation mapping technique (Kaspi et al. 2000); 3) relation between BH mass and 
its host galaxy magnitude $\log M_{\rm BH}/M_{\odot}=-0.5M_{\rm R}-2.96$ (McLure \& Dunlop 
2002); 4) relation between the dispersion velocity and BH mass 
$M_{\rm BH}=1.35\times 10^8(\sigma/200)^{4.02}$ (Tremaine et al. 2002). The black hole masses 
in Table 1 are taken from Woo \& Urry (2002) and Vestergaard (2002), otherwise we use the 
empirical 

\figurenum{1}
\centerline{\includegraphics[angle=-90,width=8.8cm]{f1.eps}}
\figcaption{\footnotesize
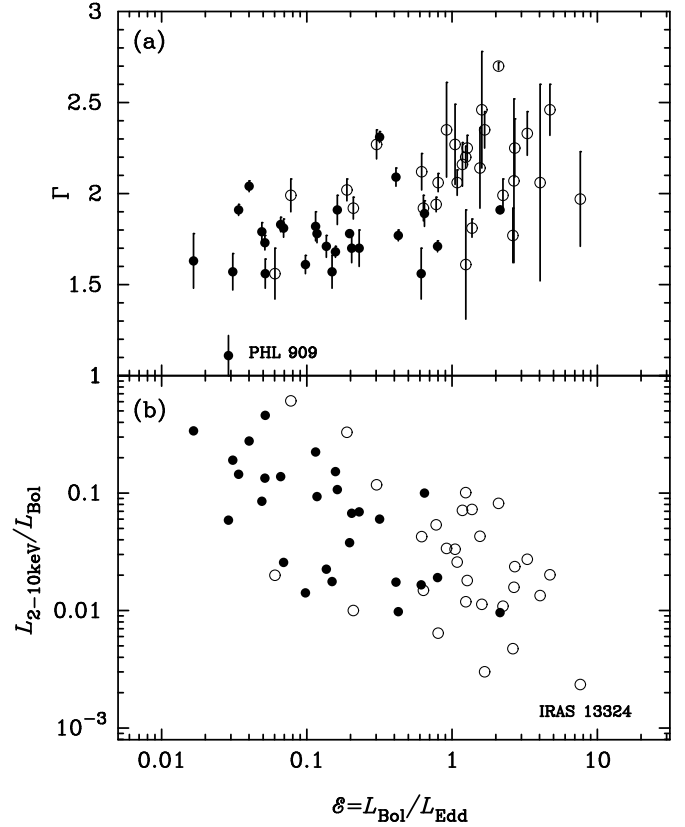
The plots of hard X-ray spectrum index and
the ratio of {\em ASCA} luminosity to bolometric luminosity versus the Eddington ratio.
The open and filled circles represent the narrow line Seyfert 1 galaxies
(FWHM H$\beta<2000$km/s)  and the  broad Seyfert 1 galaxies, respectively.
The object IRAS 13324-3809 is a NLS1 with FWHM(H$\beta)=620$km/s and 2-10keV 
luminosity $10^{43}$erg/s  much fainter than its 
$M_{\rm B}=-24.2$ (V\'eron-Cetty \& V\'eron's catalog), with $m_{\rm BH}=10^{6.95}$, 
the ratio $L_{\rm 2-10keV}/L_{\rm Edd}\sim 10^{-3}$ and 
$L_{\rm Bol}/L_{\rm Edd}\sim 10^{0.88}$. 
}
\label{fig1}
\vglue 0.5cm

\noindent
reverberation method to estimate $M_{\rm BH}$. $L_{5100\AA}$ is obtained from the 
absolute $M_{\rm B}$ given in V\'eron-Cetty \& V\'eron's quasar catalog (11th edition) through 
an extrapolation of a power law spectrum in optical band as $F_{\nu}\propto \nu^{-0.5}$. 
The bolometric luminosities $L_{\rm Bol}$ are taken from Woo \& Urry 
(2002) and Grupe et al. (2003), otherwise they are estimated via $L_{\rm Bol}=9L_{5100}$ 
(Kaspi et al. 2000). We use the Hubble constant $H_0=75$km~s$^{-1}$~Mpc$^{-1}$ 
and deceleration factor $q_0=0.5$ in this paper.

\def\ascs{{a_s}/{c_s}}
\def\trphi{t_{r\phi}}
\def\casea{$~\trphi=-\al P_{\rm tot}$}
\def\caseb{$~\trphi=-\al P_{\rm gas}$}
\def\casec{$~\trphi=-\al P_{\rm rad}$}
\def\cased{$~\trphi=-\al \sqrt{P_{\rm rad}P_{\rm tot}}$}
\def\casee{$~\trphi=-\al \sqrt{P_{\rm gas}P_{\rm tot}}$}
\def\casef{$~\trphi=-\al \sqrt{P_{\rm gas}P_{\rm rad}}$}
\def\caseg{$~\trphi=-\al P_{\rm gas}^{1/4}$}
\def\fzero{$f\rightarrow 0$}
\def\fconst{$f\rightarrow$ const.}
\def\fcasef{$\propto \al^{1/2}\sqrt{\ascs}\left[1-(\ascs)^2\right]^{1/4}$}
\def\fcasee{$\propto \al^{1/2} \sqrt{\ascs}$}
\def\fcased{$\propto \al^{1/2}\left[1-(\ascs)^2\right]^{1/4}$}
\def\fcasec{$\propto \al^{1/2}\left[1-(\ascs)^2\right]^{1/2}$}

\def\dotmg{\dot{m}^{-\gamma}}
\def\dotmgg{\dot{m}^{-\gamma/2}}
 
\def\pradgas{$P_{\rm rad}\gg P_{\rm gas}$}
\def\pgasrad{$P_{\rm gas}\gg P_{\rm rad}$}
\def\al{\alpha}
 
\begin{table*}[t]
\begin{center}
\footnotesize
\centerline{\sc Table 2. Magnetic Turbulence as Viscosity and Properties of Hot Corona}
\vskip 0.1cm
\begin{tabular}{lllllc}\hline \hline
viscosity stress &  References &factor $f$ & \pradgas &\pgasrad &Comments \\ \hline
1)\casea &SS73    &$\propto \al^{1/2}$       &$f=$ const. & $f=$ const.&  $\times$\\
2)\caseb &SR84    &$\propto \al^{1/2}\ascs$  &$f\propto \dotmg$& \fconst & $\surd$\\
3)\casec &WL91     &\fcasec &$f\propto \al^{1/2}$&\fzero & $\times$\\
4)\cased &S90    &\fcased &$f\propto \al^{1/2}$& \fzero & $\times$\\
5)\casee &TL84, B85, MF02&\fcasee &$f\propto \dotmgg$&\fconst &$\times$ \\
6)\casef &LN  89    &\fcasef&$f\propto \al^{1/2}\dotmgg$&\fzero&$\times$ \\ \hline
\end{tabular}
\vskip 2pt
\parbox{5.8in}
{\baselineskip 9pt
\indent
{\sc References:--}
B85: Burm (1985); 
MF02: Merloni \& Fabian (2002). 
LN89: Laor \& Netzer (1989); 
S90:  Szuszkiewicz (1990); 
SS73: Shakura \& Sunyaev (1973); 
SR84: Stella \& Rosner (1984); 
TL84: Taam \& Lin (1984);
WL92: Wandel \& Liang (1992).
Here $\gamma$ is a positive constant, and $a_s=(P_{\rm gas}/\rho)^{1/2}$
the gas sound speed.}
\end{center}
\end{table*}
\normalsize

Fig 1 shows the plot of photon index, $\Gamma$, and fraction of X-ray emission,
${\cal F}_{\rm X}\equiv L_{\rm 2-10keV}/L_{\rm Bol}$, versus the Eddington ratio, 
${\cal E}\equiv L_{\rm Bol}/L_{\rm Edd}$. Except for PHL 909, the X-ray photon indices 
almost keep a constant for BLS1s with a mean value of $\langle \Gamma\rangle=1.78\pm 0.17$. 
There is a trend for NLS1s: the higher is accretion rate, the steeper is the X-ray spectrum
as $\Gamma\propto \log {\cal E}^{0.18}$.
We find the Pearson's correlation coefficient of $\Gamma- {\cal E}$ is 0.37 for NLS1s.  
The difference of ${\cal E} - \Gamma$ relations in BLS1s and NLS1s show an interesting
evidence for the bi-modal accretion in Seyfert 1 galaxies. However, 
it is interesting to note that there is a strong correlation of $\Gamma$ and the Eddington 
ratio ${\cal E}$ for the entire sample,
\begin{equation}
\Gamma=(2.05\pm 0.04)+(0.26\pm 0.05)\log{\cal E},
\end{equation} 
with the Pearson's coefficient $r=0.61$ and probability $p=5.4\times 10^{-7}$.
We use the least square method of multivariate regression and find a very
strong correlation 
$\log L_{\rm X}=(0.36\pm 0.09)\log L_{\rm Bol}+(0.47\pm 0.09)\log L_{\rm Edd}+(6.13\pm 3.57)$ 
with coefficient $r=0.83$ and $F=58.7$ of the $F$-test (much larger than 
$F=3.2$ at $99.9\%$ level of confidence). This correlation 
can be rewritten as
\begin{equation}
\log {\cal F}_{\rm X}=-0.64\log {\cal E}-0.17\log L_{\rm Edd}+6.13,
\end{equation}
from which an interesting conclusion follows:  the hot corona is mainly controlled 
by the Eddington ratio. Fig 1b. shows this correlation.

The correlation shows that the corona becomes dramatically 
weak with increases of the Eddington ratio. 
While the Eddington ratio is low, a larger fraction of the energy 
will be released in the hot corona. A naive expectation is that
the magnetic stress responsible for the 
transportation of angular momentum increases with the accretion rates.  
Otherwise the accretion would
be halted because of the inefficient transportation. 
However, this does not agree with the observed correlation, which indicates that 
the dissipation in the cold disk (not in the hot corona) becomes more efficient 
with increases of the accretion rates. 
Successful models of the hot corona should address this issue explicitly.

The correlation of ${\cal E}-\Gamma$ shows that the hard X-ray 
spectrum becomes harder with increases of ${\cal E}$, confirming the results
of Lu \& Yu (1999) and Gierlinski \& Done (2004). This is usually explained by
that the seed photons efficiently cool the hot corona with increases of $\dot{m}$. 
However, this explanation could be replaced by that the hot corona really becomes 
weak since ${\cal F}_{\rm X}\propto {\cal E}^{-0.64}$. Indeed this agrees with that 
the broad line Seyfert 1 galaxies have strong reprocessing of hard X-ray, 
such as in NGC 5548 (Clavel et al. 1992; Chiang et al. 2000), evidenced by that
optical and UV variations follow the hard X-ray. In NSL1s the reprocessing is
rather weak, such as Akn 564 (Shemmer et al. 2001), NGC 4051 (Done et al. 1990) and
IRAS 13224-3809 (Young et al. 1999).

\section{Maxwell Stress and Hot Corona}
Assuming that the energy transportation from the cold 
disk to the hot corona is by buoyancy driving the magnetic tubes across extent of 
the disk, Merloni \& Fabian (2002) proposed the fraction of the energy transported 
by magnetic buoyancy to be $f=P_{\rm mag}\upsilon_{\rm P}/{Q_+}$, where $\upsilon_{\rm P}$ 
is the transporting velocity, $P_{\rm mag}$ the magnetic pressure and the dissipated 
energy $Q_+=-\frac{3}{2}c_st_{r\phi}$ via viscosity stress $t_{r\phi}$. 
We have
\begin{equation}
f=\frac{2b}{3k_0}\frac{\upsilon_{\rm A}}{c_s}
 =\frac{2^{3/2}b}{3k_0}\left(\frac{P_{\rm mag}}{P_{\rm tot}}\right)^{1/2},
\end{equation}
where $t_{r\phi}=-k_0P_{\rm mag}$, $\upsilon_{\rm A}=B/(4\pi \rho)^{1/2}$ Alfv\'en 
velocity and the constant $b=\upsilon_{\rm P}/\upsilon_{\rm A}$.
Here $P_{\rm tot}=P_{\rm rad}+P_{\rm gas}$ and $c_s^2=P_{\rm tot}/\rho$. 

\figurenum{2}
\centerline{\includegraphics[angle=-90,width=8.8cm]{f2.eps}}
\figcaption{\footnotesize
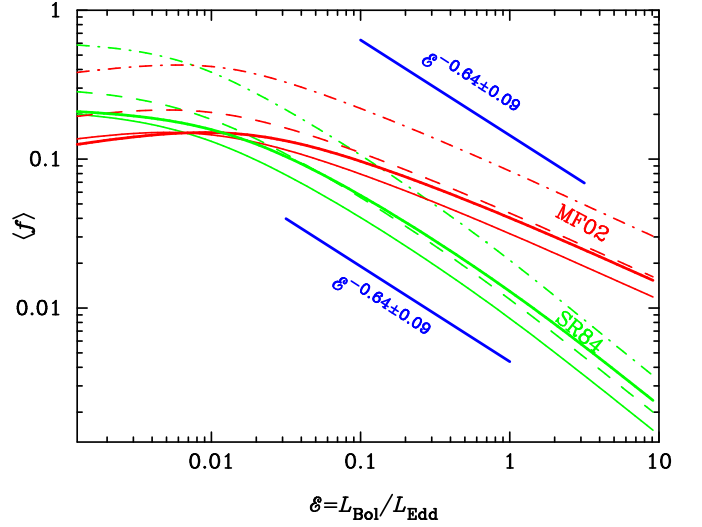
The plot of factor $f$ and the Eddington ratio in theoretical models. The lines
labeled by MF02 and SR84 are results of Merloni \& Fabian (2002) and Stella \& Rosner 
(1984), respectively. We take $2^{3/2}b/3k_0=1$ and $b=1$. 
The solid, dashed, and dot-dash-dot-dash 
lines are results for $\alpha=0.05, 0.1, 0.4$ for 
$m_{\rm BH}=10^8$. The two thick lines are for $m_{\rm BH}=10^6$ and $\alpha=0.05$.
${\cal E}=\eta \dot{m}$ is used, where the accretion 
efficiency $\eta=0.1$.}
\vglue 0.5cm
\label{fig2}

According to eq.(4), Table 2 gives the properties of the factor $f$ for 
six distinct types of the magnetic stresses based on different assumptions.
Comments are given in Table 2 based on the correlation 
(eq. 3). Cases 1), 3), 4) and 6) predict properties
inconsistent  with the correlations in Figure 1b, thus they could be ruled out.
Only Cases 2) and 5) are possible since they provide characters of the hot corona
consistent with the correlation.

Specifying the magnetic stress, we can solve the structure of disk and the factor $f$ at
each radius formulated in Svensson \& Zdziarski (1994) as described by
Merloni \& Fabian (2002) and Merloni (2003). We define the radial-averaged fraction $\pf$,
\begin{equation}
\pf=\frac{\int_{3}^{\infty}f(r)Q_{+}(r)rdr}{\int_{3}^{\infty}Q_{+}(r)rdr},
\end{equation}
so that we can compare with the observation, if the magnetic stress is assumed.

In Case 2) the magnetic pressure $t_{r\phi}=-\alpha P_{\rm gas}$ is originally suggested 
by Stella \& Rosner (1984), who argued the buoyancy controls the growth of the magnetic 
field. The magnetic stress is simply scaled by the gas pressure. Such a 
stress can stabilize the thermal instability of the cold disk (Svensson \& Zdziarski
1994). On the other hand,
Ichimaru (1975) argued that the growth of the magnetic field may be controlled by
the reconnection and gave a magnetic pressure tensor
$t_{r\phi}=-\alpha \sqrt{P_{\rm gas}P_{\rm tot}}$ (Case 5). This stress has been 
extensively studied by many authors (Burm 1985, Szuszkiewicz 1990). Recently 
Merloni \& Fabian (2002) and Merloni (2003) derived this simple formulation based 
on analytical work (Blaes \& Socrates 2001) and 3-D MHD numerical simulation  
(Turner et al. 2002) of magneto-rotational-instability in radiative accretion disks. 

Figure 2. shows the results of $\pf$ for the magnetic stresses 
$t_{r\phi}=-\alpha P_{\rm gas}$ and $t_{r\phi}=-\alpha\sqrt{P_{\rm gas}P_{\rm tot}}$. 
We find that $\pf$ is not sensitive to the black hole mass.
More interesting thing in Figure 2 is that the slope of ${\cal E}-\pf$ (for ${\cal E}\ge 0.05$)
is insensitive to both the black hole mass and the viscosity $\alpha$, but only sensitive to
the magnetic stress. These properties of the theoretical relation of ${\cal E}-\pf$  
allow us to test the working magnetic stress in disk.
We have $\pf\propto {\cal E}^{-0.44}$ for the Merloni \& Fabian's model 
whereas $\pf\propto {\cal E}^{-0.77}$ for the Stella \& Rosner's. MF02 predicts a flatter slope 
than the observation. SR84 model has a steeper slope, but it
matches the observational correlation $\pf \propto {\cal E}^{-0.64\pm 0.09}$ within 
the uncertainties in the present sample. 

A more fiducial test on
the magnetic stress requires the accurate estimations of the black hole
masses, bolometric luminosity and the total hard X-ray luminosity from observations.
The near future observations of {\em INTEGRAL} 
can provide entire hard X-ray spectra of a large sample so as to test the disk-corona
models. The future 3-D MHD numerical simulations including the hot corona should be tested 
by the plot of ${\cal E}-L_{\rm X}/L_{\rm Bol}$.

\section{Conclusions}
With the estimations of the black hole masses in radio-quiet AGNs, 
we find the hard X-ray spectrum index and $L_{\rm 2-10keV}/L_{\rm Bol}$ strongly 
correlate with the Eddington ratio. These correlations
directly indicate that the fraction $f$ is controlled by the Eddington ratio.
Such a correlation can be explained by the magnetic turbulence that plays two roles 
in transporting angular momentum and forming the hot corona.
The correlation ${\cal F}_{\rm X}\propto {\cal E}^{-0.64\pm 0.09}$ favors
the magnetic stress $t_{r\phi}\propto P_{\rm gas}$ 
in the present sample.

\acknowledgements
This research is supported by Grant for Distinguished Young Scientists from 
NSFC, 973 project and a Grant-in-Aid of the 21st Century COE "Center for Diversity and 
Universality in Physics". 

\vglue -0.2cm

\clearpage

\clearpage

\clearpage

\clearpage


\end{document}